# Laboratory Simulations of Haze Formation in the Atmospheres of super-Earths and mini-Neptunes: Particle Color and Size Distribution

Chao He[1], Sarah M. Hörst[1], Nikole K. Lewis[1,2], Xinting Yu[1], Julianne I. Moses[3], Eliza M.-R. Kempton[4], Patricia McGuiggan[5], Caroline V. Morley[6], Jeff A. Valenti[2], and Véronique Vuitton[7]

[1] Department of Earth and Planetary Sciences, Johns Hopkins University, Baltimore, MD, USA che13@jhu.edu

[2] Space Telescope Science Institute, Baltimore, MD, USA

[3] Space Science Institute, Boulder, CO, USA

[4] Grinnell College, Grinnell, IA, USA

[5] Department of Materials Science and Engineering, Johns Hopkins University, Baltimore, MD, USA

[6] Harvard University, Cambridge, MA, USA

[7] Univ. Grenoble Alpes, CNRS, IPAG, 38000 Grenoble, France

## Abstract:

Super-Earths and mini-Neptunes are the most abundant types of planets among the ~3500 confirmed exoplanets, and are expected to exhibit a wide variety of atmospheric compositions. Recent transmission spectra of Super-Earths and mini-Neptunes have demonstrated the possibility that exoplanets have haze/cloud layers at high altitudes in their atmospheres. However, the compositions, size distributions, and optical properties of these particles in exoplanet atmospheres are poorly understood. Here, we present the results of experimental laboratory investigations of photochemical haze formation within a range of planetary atmospheric conditions, as well as observations of the color and size of produced haze particles. We find that atmospheric temperature and metallicity strongly affect particle color and size, thus altering the particles' optical properties (e.g., absorptivity, scattering, etc.); on a larger scale, this affects the atmospheric and surface temperature of the exoplanets, and their potential habitability. Our results provide constraints on haze formation and particle properties that can serve as critical inputs for exoplanet atmosphere modeling, and guide future observations of super-Earths and mini-Neptunes with the Transiting Exoplanet Survey Satellite (TESS), the James Webb Space Telescope (JWST), and the Wide-Field Infrared Survey Telescope (WFIRST).

1. INTRODUCTION

The Kepler mission has shown that the most numerous types of planets are super-Earths (1.25 $R_{Earth}$ < $R_p$ < 2.0 $R_{Earth}$) and mini-Neptunes (2.0 $R_{Earth}$ < $R_p$ < 4.0 $R_{Earth}$) among the ~3500 confirmed exoplanets (e.g., Borucki et al. 2011, Fressin et al. 2013). The Transiting Exoplanet Survey Satellite (TESS) mission will further increase the number of super-Earths and mini-Neptunes, including a sample of rocky worlds in the habitable zones of their host stars. These types of exoplanets are expected to exhibit a wide variety of atmospheric compositions (e.g., Elkins-Tanton & Seager 2008, Miller-Ricci et al. 2009, Schaefer et al. 2012, Moses et al. 2013, Hu & Seager 2014, Venot et al. 2015, Ito et al. 2015), and TESS will seek out targets for atmospheric characterization by the James Webb Space Telescope (JWST), as well as other large ground-based and space-based telescopes in the future. Clouds and hazes are likely to be present in exoplanet atmospheres, as they are ubiquitous in the atmospheres of solar system worlds. Recent transmission spectra have demonstrated that clouds and/or hazes could play a significant role in atmospheres of Neptune-sized GJ436b (Knutson et al. 2014a), and super-Earth/mini-Neptune sized planets GJ1214b (Kreidberg et al. 2014) and HD97658b (Knutson et al. 2014b). Condensate cloud and photochemical haze particles affect the chemistry, dynamics, and radiation flux in planetary atmospheres, and can therefore influence surface temperature and habitability. However, little is known about the composition, size distributions, and optical properties of the particles that may form in exoplanet atmospheres, which largely exist outside the size and temperature regimes of solar system planets. Laboratory studies of exoplanet haze particles are essential to interpreting future spectroscopic observations and characterizing the atmospheres of these worlds. We have reported the formation of hazes in laboratory simulations of the atmospheres expected for super-Earths and mini-Neptunes (Hörst et al. 2018), and here we present the haze colors and the particle sizes measured with Atomic Force Microscopy (AFM). Besides dramatically different production rates and haze colors, the AFM images show that the particle size varies with initial gas composition.

Particle size affects the scattering and absorption of light, and thus impacts the temperature structure of exoplanet atmospheres as well as their observed spectra. Remote and in situ measurements have provided useful information on particle sizes in planetary atmospheres of solar system bodies; for example, the Cassini–Huygens and New Horizons missions have put important constraints on the haze particle sizes of Titan and Pluto (Tomasko et al. 2005 & 2008, Gladstone et al. 2016). However, very little is known about particle sizes in exoplanet atmospheres. Laboratory study of exoplanet haze analogs could place some constraints on the particle sizes. Such studies of Titan haze analogs (tholins) have helped understand the size and growth of the haze particles in Titan's atmosphere. The particle sizes of Titan tholins prepared in laboratories have been measured through several different techniques, including Scanning Mobility Particle Sizer (SMPS, see e.g., Trainer et al. 2006; Hörst & Tolbert 2013 & 2014), Scanning Electron Microscopy (SEM, see e.g., Szopa et al. 2006, Hadamcik et al. 2009, Carrasco et al. 2009, Sciamma-O'Brien et al. 2017), Transmission Electron Microscopy (TEM, see e.g., Trainer et al. 2006, Quirico et al. 2008, Curtis et al. 2008), and AFM (Li et al. 2010, Hasenkopf et al. 2011). SEM and TEM usually require a conductive coating on the sample and could also cause thermal damage to the sample. We use AFM in this study because it can image the original sample surface without coating an additional layer or exposing the sample to electron radiation.

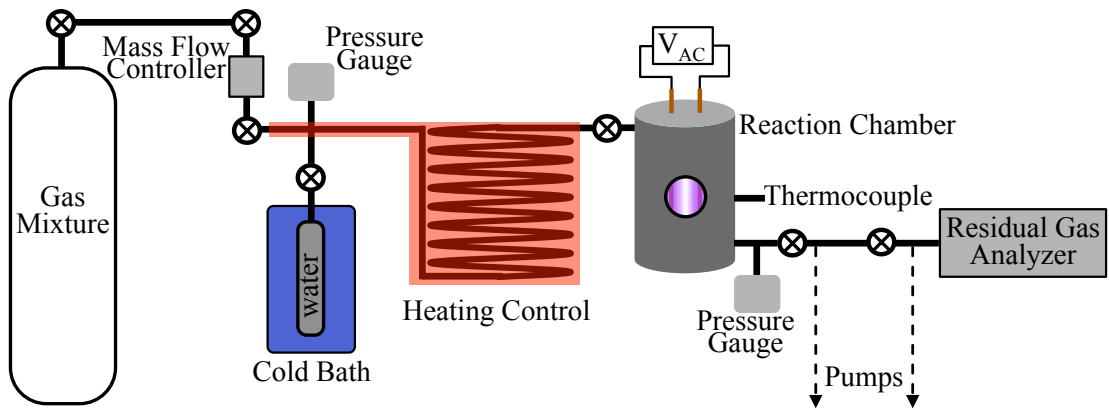

**Figure 1.** Schematic of the experimental setup used for this work. Gas mixtures for each experiment excluding water vapor (and ammonia for the 300 K-100× metallicity experiment) are prepared in respective ratios as listed in Table 1 and mix overnight in the mixing cylinder. Water vapor (and ammonia for the 300 K-100× metallicity experiment) is supplied by water (or ammonia water solution) at the desired temperature maintained by a dry ice/methanol/water cold bath. The prepared gas mixture and water vapor (and ammonia for the 300 K-100× metallicity experiment) flow through a heating coil where the gases are heated to the experimental

temperature, and into the reaction chamber where they are exposed to an AC glow discharge, initiating chemical processes that lead to the formation of new gas phase products and particles.

Table 1. Initial gas mixtures used in the experiments. Our experimental phase space spans 100 to 10,000x solar metallicity and temperatures ranging from 300 to 600 K.

|  | 100× | 1000× | 10000× |
|---|---|---|---|
| 600 K | 72.0% $H_2$<br>6.3% $H_2O$<br>3.4% $CH_4$<br>18.3% He | 42.0% $H_2$<br>20.0% $CO_2$<br>16.0% $H_2O$<br>5.1% $N_2$<br>1.9% CO<br>1.7% $CH_4$<br>13.3% He | 66.0% $CO_2$<br>12.0% $N_2$<br>8.6% $H_2$<br>5.9% $H_2O$<br>3.4% CO<br>4.1% He |
| 400 K | 70.0% $H_2$<br>8.3% $H_2O$<br>4.5% $CH_4$<br>17.2% He | 56.0% $H_2O$<br>11.0% $CH_4$<br>10.0% $CO_2$<br>6.4% $N_2$<br>1.9% $H_2$<br>14.7% He | 67.0% $CO_2$<br>15.0% $H_2O$<br>13.0% $N_2$<br>5.0% He |
| 300 K | 68.6% $H_2$<br>8.4% $H_2O$<br>4.5% $CH_4$<br>1.2% $NH_3$<br>17.3% He | 66.0% $H_2O$<br>6.6% $CH_4$<br>6.5% $N_2$<br>4.9% $CO_2$<br>16.0% He | 67.3% $CO_2$<br>15.6% $H_2O$<br>13.0% $N_2$<br>4.1% He |

## 2. MATERIALS AND EXPERIMENTAL METHODS

### 2.1. Haze Production Setup

Figure 1 shows a schematic of Planetary Haze Research (PHAZER) experimental setup at Johns Hopkins University (He et al. 2017, Hörst et al. 2018). The initial gas mixtures for our experiments are calculated from the chemical equilibrium models of Moses et al. (2013), who examined the possible thermochemistry and photochemistry in the atmospheres of Neptune-sized and sub-Neptune-sized exoplanets. Most of the known super-Earth and mini-Neptune size planets known to date were discovered by the Kepler mission (Borucki et al. 2011, Fressin et al. 2013). However, follow-up of Kepler targets to determine planetary mass and atmospheric characterization has been limited due to the brightness of their host stars. The TESS mission (Ricker et al., 2014) will target stars that are 10-100 times brighter than those targeted with the Kepler mission and yield a large sample of super-Earth and mini-Neptune class planets with equilibrium temperatures (Teq) < 1000 K (Sullivan et al., 2015). This planetary size ($R_p < 4R_{Earth}$) and temperature (Teq <1000 K) phase space has been recently explored by exoplanet atmospheric characterization efforts for a handful of planets such as GJ 1214b, HD 97658b, GJ 436b, GJ 3470b, which all indicate that aerosols are shaping the planetary spectrum (e.g. Kreidberg et al. 2014; Knutson et al. 2014a, 2014b; Dragomir et al. 2015). Theoretical studies predict a diverse range of possibilities for the composition of atmospheric envelopes of mini-Neptunes and super-Earths (e.g. Fortney et al. 2013, Wolfgang & Lopez 2015, Howe & Burrows 2015). Broadly speaking, we expect that our hydrogen-rich gas mixtures are more relevant to larger mini-Neptune class planets and carbon dioxide-rich gas mixtures are more relevant to smaller super-Earth class planets, with water-rich gas mixtures being relevant to both classes of planets. However, further observational, theoretical, and laboratory work will be required to confirm or refute the potential diversity in atmospheric compositions for these classes of planets that do not exist in our solar system. The equilibrium calculations relevant to this investigation were performed at conditions of 300, 400, and 600 K at 1 mbar for 100×, 1000×, and 10000× solar metallicity. Smaller planets are less efficient at accreting the H- and He-rich gas from the protoplanetary disk, and the lighter elements have a greater chance of escaping from smaller planets, which enhances the overall metallicity of planets Neptune-sized and smaller. As an example, the O/H metallicity on Uranus and Neptune is estimated to be <160 and 540 times solar, respectively (Cavalie et al. 2017). Various evolutionary processes could drive the relative abundances of the heavy elements away from solar, but these high-metallicity chemical-equilibrium models provide a reasonable starting point for our study. As listed in Table 1, we included gases with a calculated abundance of 1% or higher to maintain a manageable level of experimental complexity; this resulted in no nitrogen-bearing species in two cases and the exclusion of sulfur-bearing species. Sulfur-bearing species, like $H_2S$, could be important for haze formation (Gao et al. 2017) and will be an avenue of future work. The pressure, temperature, and gas compositions used in the experiments are self-consistent based on the model calculations.

Gas mixtures for each experiment, excluding water vapor (and ammonia for the 300 K-100× metallicity experiment), were prepared in the ratios as listed in Table 1, respectively, and mixed overnight in the mixing cylinder. High purity gases purchased from Airgas were used to prepare the mixture ($H_2$-99.9999%, He-99.9995%, $N_2$-99.9997%, $CH_4$-99.999%, $CO_2$-99.999%, CO-99.99%). A liquid nitrogen cold trap was used to remove known contaminants from CO before mixing it with the other gases (see e.g., Hörst & Tolbert 2014; He et al. 2017). The total flow rate is 10 standard cubic centimeters per minute (sccm), thus we flow the gas mixture (without water and ammonia) at a proportional rate of 10 sccm based on the ratio in Table 1, and measure the pressure ($P_1$) at the same time. The pressure ($P_2$) of water vapor is then calculated ($P_2/P_1=$ the mixing ratio of water to other gases as in Table 1). For instance, in the 600 K-100× metallicity experiment, $P_1$ is 4.90 Torr when the flow rate of the gas mixture is 9.37 sccm, thus $P_2$ is calculated to be 0.33 Torr. The gas mixture is controlled by a mass flow controller (MKS Instrument, GM50A) and water vapor is supplied by HPLC water (Fisher Chemical) at desired temperature maintained by a dry ice/methanol/water cold bath. The temperature of the cold bath is adjusted by varying the methanol/water ratio. For the 300 K-100× metallicity experiment, a 0.5 wt% ammonia water solution (prepared with HPLC water and 28% $NH_3$ in $H_2O$, EMD) at 246 K provides water and ammonia vapor (Field & Combs, 2002). The prepared gas mixture and water vapor (and ammonia for the 300 K-100× metallicity experiment) flow through a 15-meter stainless steel heating coil where the gases are heated to the experimental temperature (600 K, 400 K, or 300 K), and then into a stainless-steel reaction chamber where they are exposed to an AC glow discharge, initiating chemical processes that lead to the formation of new gas phase products and solid particles. The plasma produced by the AC glow discharge is not analogous to any specific process in planetary upper atmospheres; we use it as a proxy for energetic processes occurring in planetary upper atmospheres, because it is sufficiently energetic to directly dissociate very stable molecules such as $N_2$ or CO. AC glow discharge, as a cold plasma source, does not alter the neutral gas temperature significantly, and is often used as an analog for the relatively energetic environment of planetary upper atmospheres (Cable et al. 2012). During 72 hr of discharge flow, produced solid particles (tholin) are deposited on the wall of the reaction chamber and quartz substrate discs (purchased from Ted Pella, Inc., made from high quality fused quartz and optical-grade clear polished on both sides) at the bottom of the chamber. We ran our previous Titan experiments for 72 hr to get sufficient samples (He et al. 2017), thus we followed the same procedure here for comparison. The chamber is further kept under vacuum for 48 hr to remove the volatile components, and then transferred to a dry (<0.1 ppm $H_2O$), oxygen free (<0.1 ppm $O_2$) $N_2$ glove box (Inert Technology Inc., I-lab 2GB) where the tholin films on quartz discs and tholin powders from the wall are collected under the dry $N_2$ atmosphere. Both the powders and films are kept in the glove box and wrapped in foil to avoid exposure to air and light, respectively.

*2.2. Atomic Force Microscopy (AFM) Measurement*

The surface morphology of the tholin films is examined using a Bruker Dimension 3100 atomic force microscope (Bruker Nano, Santa Barbara, CA). A silicon probe (Tap300-G, Ted Pella, Inc.) is used, and its tip radius is less than 10 nm and conical angle at the apex is less than 20°. The AFM images are acquired by scanning the sample under ambient laboratory conditions (298 K) at a scan rate of 1.5 Hz using tapping mode.

3. RESULTS AND DISCUSSION

*3.1. Tholin Films and Production Rate*

Figure 2 shows pictures of the tholin films deposited on quartz discs (1" diameter) in visible light. The color of the films varies significantly as a function of metallicity and temperature. For comparison, we include the picture of Titan tholin film on quartz disc from our standard Titan experiment (5% $CH_4$ in 95% $N_2$) using the identical setup (He et al. 2017). The production rates of the haze particles produced from the nine experiments (Hörst et al. 2017) and our previous Titan experiment (He et al. 2017) are also included in Figure 2. Among the nine films in Figure 2, those from the 1000× metallicity at 400 K and 300 K experiments are dark gold in color, and the one at 300 K is slightly lighter than that at 400 K. The color of those two films looks very similar to that of our standard Titan experiment. As we reported previously (Hörst et al. 2018), the production rate of these two cases is also compatible to our standard Titan experiment (~10 mg/hr vs. 7.4 mg/hr). The production rate and color may suggest the similarities in the chemical processes leading to the particle formation and the resulting compositions, so further work is needed to understand the complex chemical systems.

For 100× metallicity, the film from 600 K experiment is olive green, while the one from 400 K looks chocolate brown. In these two cases, the production rate is relatively low (0.04 mg/hr and 0.25 mg/hr, Hörst et al. 2018), thus the relatively dark films are probably derived from dark colored particles rather than thick films. Two distinct colors suggest compositional difference for the two films. Considering the similarity of the gas mixtures in these two cases, the temperature (600 K versus 400 K) could play an important role in the atmospheric chemistry that leads to different compositions of the hazes. The films from 100× metallicity at 300 K, 1000× metallicity at 600 K, and 10000× metallicity at 600 K experiments are all light yellow in color, consistent with the low production rate in these cases (Hörst et al. 2018). Both films from 10000× metallicity at 400 K and 300 K show no obvious difference with the clear blank quartz disc, suggesting very few particles deposited on the substrate in these two cases.

The films shown in Figure 2 demonstrate large variations in particle color at different temperatures for varying metallicity. The haze particles in different colors will absorb and scatter visible light differently, and thus affect the thermal structure of the exoplanet atmospheres and the potential habitability. To better understand these effects, optical constant measurements (in a wide range of wavelengths) of these analog materials are underway.

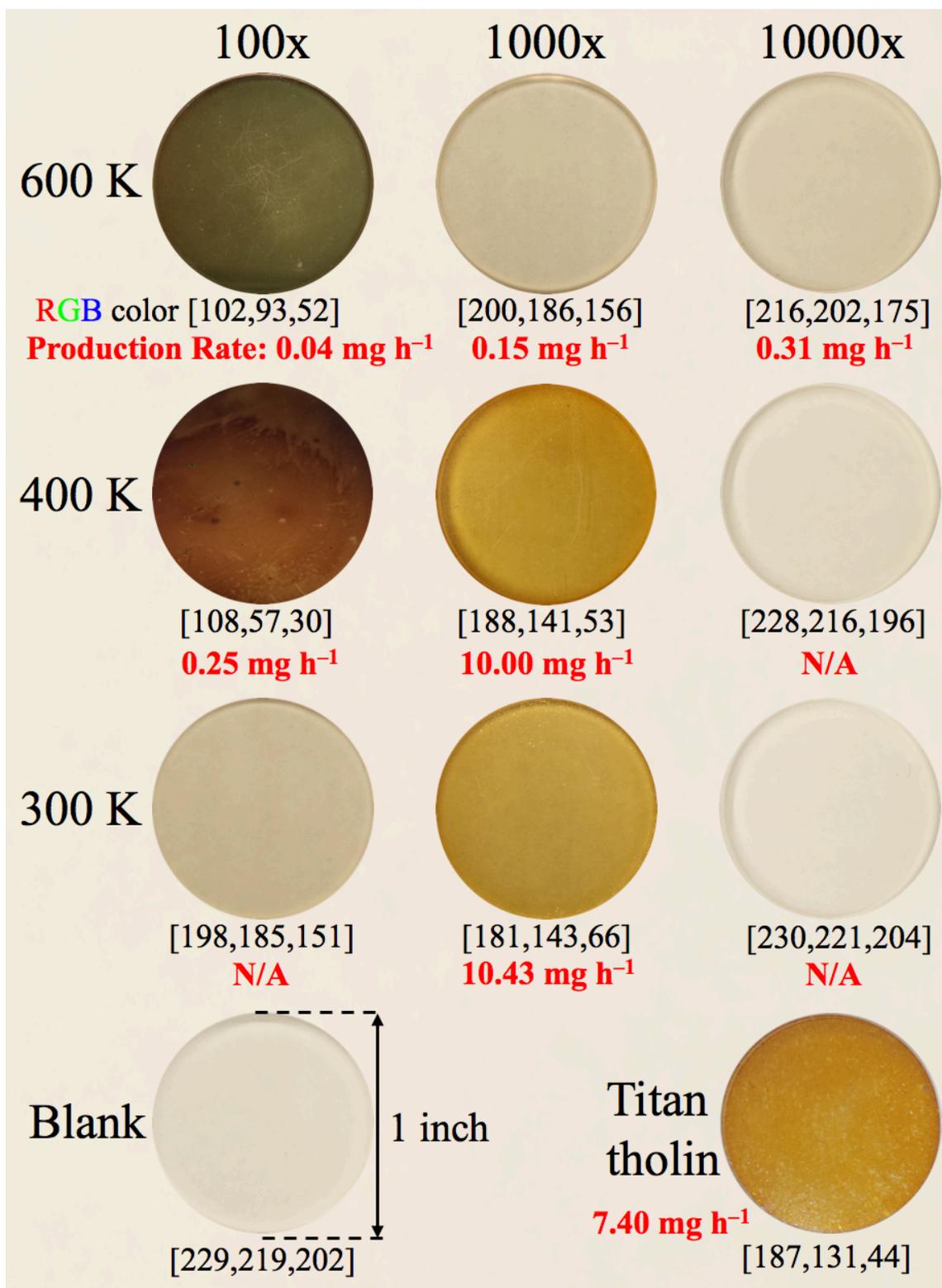

**Figure 2.** Pictures of the tholin films deposited on quartz discs [diameter: 1"(2.54 cm)]. Blank is the picture of a clean blank quartz disc, and Titan tholin is the picture of the Titan tholin film on quartz discs from our standard Titan experiment (5% $CH_4$ in 95% $N_2$). Average red, green, and blue (RGB) colors of the pictures are shown in the brackets under the pictures. Production rates (mg h$^{-1}$) the haze particles produced from the nine experiments (Hörst et al. 2018) and our previous Titan experiment (He et al. 2017) are shown under the pictures in red. N/A indicates that the solid sample was not enough to collect and weigh.

*3.2. AFM Images of the Tholin Films and the Particle Size Distribution*

Figure 3 shows AFM 3-dimensional (3D) images of the tholin films deposited on quartz discs, displaying 1 μm x 1 μm scanning area for each film. An AFM image of a clean blank quartz disc is also included in Figure 3, showing a smooth, clean surface. Compared to the clear blank disc, we observe obvious spherical particles on the discs from 9 experiments, confirming that haze particles are produced in all nine diverse atmospheres. This is particularly important for the 10000× metallicity at 400 K and 300 K experiments, because it is difficult to determine whether or not haze particles are formed by visual examination of these discs (Figure 2). Figure 3 shows that the particles sparsely spread on these two discs, indicating the relatively low production rate in these two cases. In contrast to these two cases, there are more particles produced for the 10000× metallicity at 600K experiment, consistent with the measured production rates (Hörst et al. 2018). The formation of hazes in the absence of methane suggests other carbon sources for the organics, such as CO and $CO_2$. The dissociation of CO or $CO_2$ produces atomic carbon, which can be used to build organic molecules. Previous studies have shown that reactions (irradiated by UV or plasma) beginning with the gas mixture of $CO/N_2/H_2O$ or $CO_2/N_2/H_2O$ produce a variety of organic compounds (See eg., Bar-Nun & Chang 1983, Plankensteiner et al. 2004, Cleaves et al. 2008). This result here demonstrates that there are multiple pathways for organic haze formation and that $CH_4$ is not necessarily required.

We measured the diameter based on the particle's projection on the x-y plane. This method gives an accurate result (measurement errors are less than 3 nm) in the size range we measured (Villarrubia 1997, Klapetek et al. 2011). The AFM image of the Titan tholin film in Figure 3 shows that numerous small particles (diameter 30 to 100 nm) are densely deposited on the quartz disc. Similarly, there are a large number of particles closely packed on the discs for the 100× and 1000× metallicity experiments at all three temperatures (Figure 3). The particles are of different sizes for different cases. The number of the particles is consistent with a higher production rate in these cases. For 100× metallicity cases, small particles (diameter 20 to 55 nm) are against each other tightly, while the particle sizes have wider distributions (25 nm to 130 nm) for 1000× metallicity cases. It is noteworthy that the particles appear to array into linear structures for 1000× metallicity at 400 K and 300 K, while these two cases have highest production rate (~10 mg/hr, Hörst et al. 2018). In these two cases, a great many particles are produced and deposited on the quartz discs in multiple layers. The AFM images only display the top layer of the particles. During the reactions, the particle interactions are governed by interparticle forces, such as van der Waals, hydrogen bonds, and weak polar forces. This may lead to the newly deposited particles forming the linear structures as in Figure 3, which could be similar to the self-assembly of nanoparticle chains of organic polymers due to hydrogen bonding (see e.g., Grzelczaket al. 2010, Kao et al. 2012). The size ranges of the haze particles in these two cases (30 to 90 nm in diameter) are also similar to the size range of the Titan tholin (25 to 80 nm). However, the linear structures are not observed on the Titan tholin film from our standard Titan experiment (Figure 3), which has the comparable production rate. Since water (present in the 1000× metallicity 300 K and 400 K experiments at 66% and 56%, respectively) is a polar molecule, it could contribute to the interparticle forces that caused the formation of the linear structures.

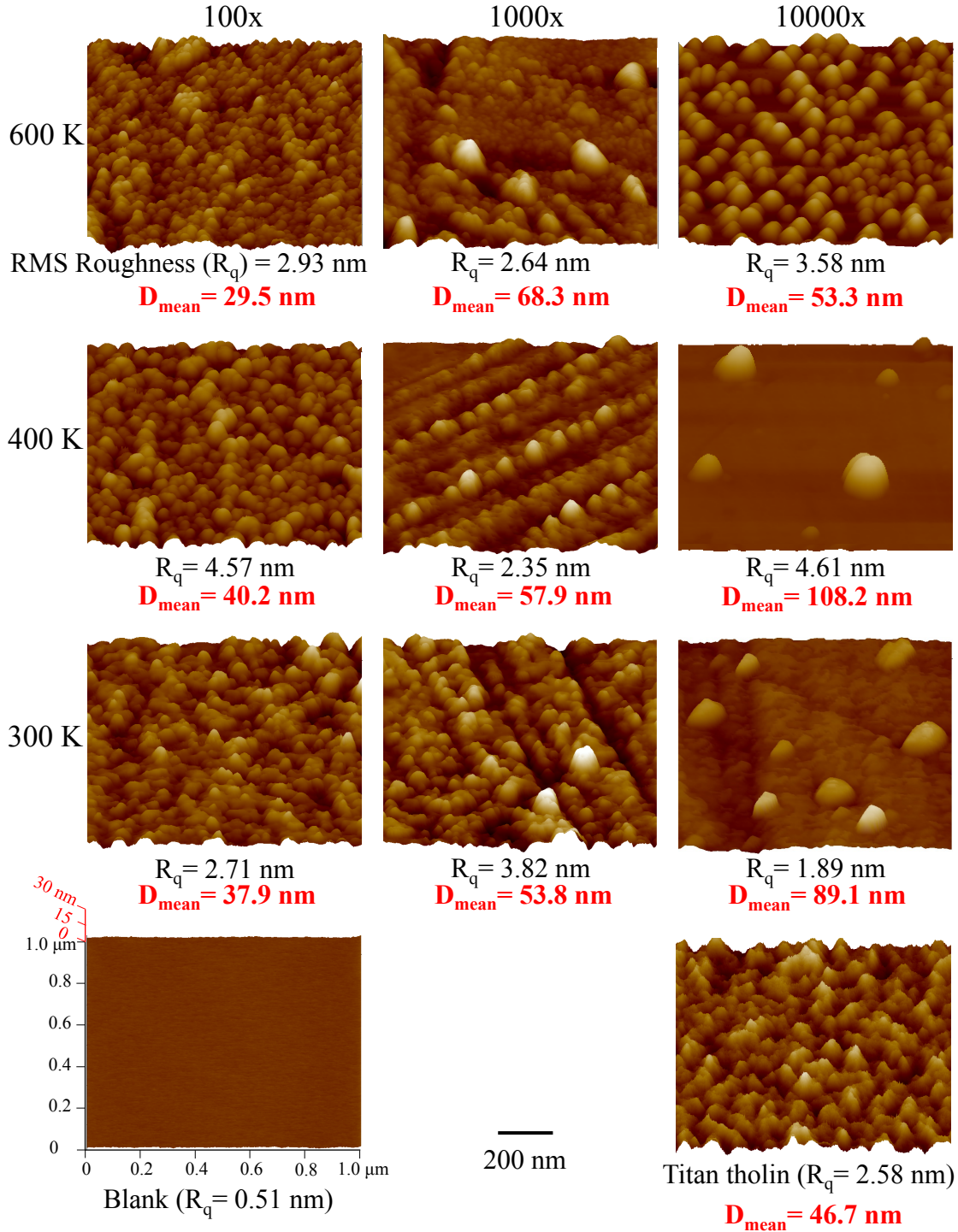

**Figure 3.** AFM 3D images of the tholin films deposited on quartz discs. Scan area is 1 μm×1 μm. Blank is the AFM image of a clean blank quartz disc, and Titan tholin is the AFM image of the Titan tholin film on quartz discs from our standard Titan experiment (5% $CH_4$ in 95% $N_2$). The height scale of the images is 30 nm as shown in the blank disc, except the one for the 400 K-10000× metallicity experiment, which has 80-nm height scale to better show the large particles. The root mean square (RMS) roughness ($R_q$) of the film (1 μm×1 μm) is shown under the images.

The roughness (< 5 nm) indicates the films are smooth. The mean diameters ($D_{mean}$, nm) of the haze particles are also shown under the images (in red).

The AFM images (Figure 3) show that the particle size varies from 20 nm to 180 nm. In order to better reveal the size distribution of the haze particles, we analyzed the particle size in a larger scanning area (10 μm x 10 μm) of each film, and plotted the percentage of particles ($N/N_{total}\times100\%$) in 1 nm bins (Figure 4). The haze particles are most uniform in size for the 100× metallicity experiments; the particles formed at 600 K are between 20 nm and 40 nm in diameter, those at 400 K are between 30 nm and 55 nm, while those at 300 K vary from 25 nm to 55 nm. The particles from the 1000× metallicity experiments have wider size distributions; the particles at 600 K are between 25 nm and 130 nm, those at 400 K are from 40 nm to 90 nm, and those at 300 K are from 30 nm to 85 nm. For the 1000×-600 K experiment, the production rate is low (0.15 mg/hr, Hörst et al. 2018), thus there are few nucleation centers in the system and the particle growth rates are slow. However, gas-solid heterogeneous reactions could happen on the particle surfaces and lead to the formation of wider range of particles. In the two high production rate cases (400 K and 300K), the bigger particles could be formed by small particle aggregation. The two high production rate cases produce sufficient numbers of small particles and they may grow into fractal aggregates, such as those observed in the detached haze layer in Titan's atmosphere (see e.g., Lavvas et al. 2009, Cours et al. 2011). The aggregates in the detached haze layer have an effective diameter around 300 nm, and are probably formed from 80 nm (diameter) monomers (Lavvas et al. 2009, Cours et al. 2011). If similar processes happened in our two high production rate cases, both the monomers (~30 nm in diameter) and the aggregates (~90 nm) are smaller than those in Titan's detached haze layer. For the 10000× metallicity experiments, the particles at 600 K are smaller and more uniform in size (30 nm to 80 nm) compared to those at 400 K and 300 K (50 nm to 180 nm). For those at 400 K and 300 K, the broader range of particle sizes could be caused by the low production rate, similar to the 1000x-600 K case. In these cases, few nucleation centers and heterogeneous reactions lead to the formation of more diverse particles. In Figure 4, several particle size distributions (the 1000× metallicity experiments at three temperatures, and the 10000× metallicity experiments at 400 K) appear to have a bimodal distribution. As we discussed above, the bimodal distribution could be caused by small particle aggregation for the high production rate cases (the 1000×-400 K and the 1000×-300 K experiments), or heterogeneous reactions with few nucleation centers for the low production rate cases (the 1000×-600 K and the 10000×-400 K experiments). Both the AFM images and the size distribution indicate that most of the particles are monomers except a few aggregates for the higher production rate cases (1000× metallicity experiments).

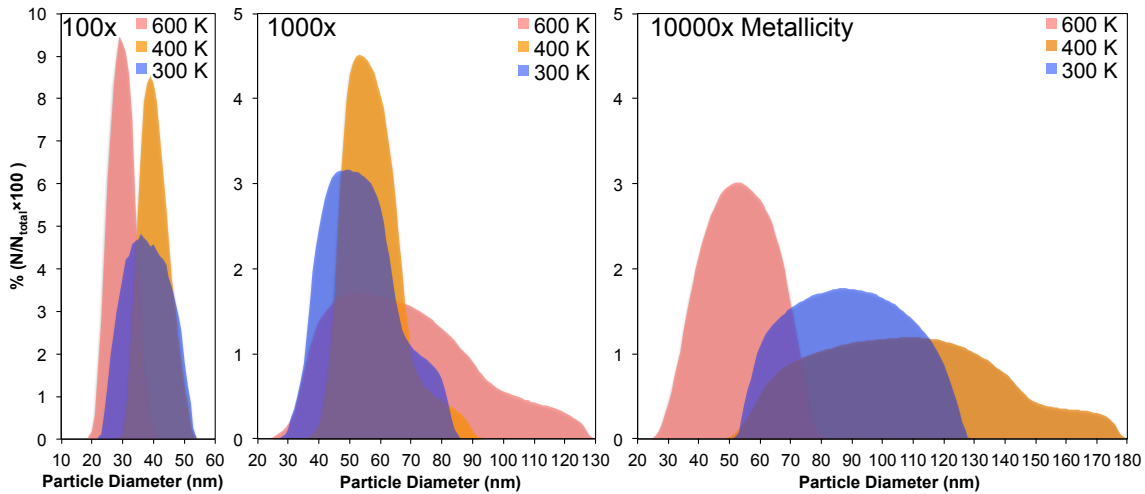

**Figure 4.** Size distribution of the tholin particles formed in the nine experiments. The haze particles are more uniform for 100× metallicity experiments (20 nm to 55 nm), while the particle sizes for the 1000× and 10000× metallicity experiments have wider distributions (25 nm to 180 nm).

Studies show that the particle size of Titan haze analogs varies with the initial gas composition (Hadamcik et al. 2009, Hörst & Tolbert 2014, Sciamma-O'Brien et al. 2017). Our experimental matrix varied in temperature and metallicity and all 9 initial gas mixtures are compositionally distinct, which may cause the differences in particle size. Future investigations are required to fully understand how the gas compositions affect the particle size. Particle size affects the interaction of haze particles with radiation, such as scattering and absorption, and thus impacts the energy budget and the temperature structure of exoplanet atmospheres. For all nine exoplanet atmospheric scenarios investigated here, the size range of the particles (20 nm to 180 nm) lie in the Rayleigh scattering regime for visible and infrared (IR) photons. In this regime, the small particles will scatter short wavelengths more efficiently. The existence of haze particles would drastically change the geometric albedo of the exoplanet in visible and near IR regions (0.4 μm to 1.0 μm) (McCullough et al. 2014, Morley et al. 2015, Sing et al. 2016, Gao et al. 2017), which could affect the detectability of directly imaged exoplanets by the Coronagraph Instrument on the Wide-Field Infrared Survey Telescope (WFIRST) (Spergel et al. 2015). The haze particles have different absorption features at different wavelengths depending on their composition. Thus, these features are helpful to identify the composition of the haze particles. However, these features could also reduce the detectability of atmospheric chemical species. To fully understand the effect of the haze particles on observations, their optical and compositional properties are required. These measurements of the analog materials are underway.

Currently, wide ranges of particle sizes, from 5 nm to 10,000 nm, are used in exoplanet atmospheric models (see e.g., Howe & Burrows 2012, Arney et al. 2016, Gao et al. 2017). However, it is difficult to constrain particle sizes by modeling. For instance, a variety of models are used to reproduce featureless transit spectrum of GJ1214b, where we see strong evidence for aerosols, but the results from different models are not consistent in the particles sizes: Morley et al. (2015) showed that photochemical hazes with a range of particle sizes (10 nm to 300 nm) can create featureless transmission spectra; while Charnay et al. (2015) found that the predicted transit spectrum matches with observations when particle radii are around 500 nm. The particles (20 nm to 180 nm) formed in our experiments fall within the size range discussed by Morley et al. (2015), suggesting that small particles are more likely in the atmosphere of GJ1214b.

Our result shows that the haze particles are more uniform for 100× metallicity experiments (20 nm to 55 nm), while the particle sizes for 1000× and 10000× metallicity experiments have wider distributions (25 nm to 180 nm). Laboratory experimental simulations cannot perfectly replicate photochemical processes that occur in exoplanet atmospheres, thus the size ranges we measured may not represent the actual haze particle sizes in exoplanet atmospheres. However, laboratory studies can provide insights into the size and growth of the haze particles in planetary atmospheres; for instance, the size of Titan tholin monomers is consistent with that of the actual haze monomers observed in Titan's atmosphere (Tomasko et al. 2005 & 2008, Trainer et al. 2006, Hörst & Tolbert 2013), and the tholin aggregates also match those observed in the detached haze layer in Titan's atmosphere (see e.g., Tomasko et al. 2008, Hadamcik et al. 2009, Cours et al. 2011). We measured the particle sizes for both monomers and aggregates (only a few for the higher production rate cases, 1000× metallicity experiments). The size distribution can be used as a guide for modeling super-Earth and mini-Neptune atmospheres with haze layers, and the size difference for different metallicities and temperatures should be considered. The size distribution of the haze particles, along with the production rate (Hörst et al. 2018), provides critical inputs for modeling the atmospheres of exoplanets, and useful data for interpreting and guiding current and future observations of super-Earths and mini-Neptunes. Further optical constant measurements and compositional analyses of these haze particles are in progress, which will guide future observations of super-Earths and mini-Neptunes atmospheres.

## 4. CONCLUSIONS

We conducted a series of laboratory experimental simulations to investigate haze formation in a range of planetary atmospheres by using PHAZER chamber (He et al. 2017), and found that haze particles are formed in all nine experiments, but the haze production rates are dramatically different for different cases (Hörst et al. 2018). We examined the color and particle size of the tholin films deposited on quartz discs. The results demonstrate large variation in particle color at different temperatures for different metallicities. The AFM images reveal that the particle size varies from 20 nm to 180 nm. The haze particles are more uniform for the 100× solar metallicity experiments (20 nm to 55 nm) while the particle sizes for the 1000× and 10000× solar metallicity experiments have wider distributions (25 nm to 180 nm). The formation of haze particles in different colors and sizes will have an impact on light absorption and scattering, thus influencing the atmospheric and surface temperature of the exoplanets and their potential habitability. The size distribution of the haze particles measured here, along with the production rate (Hörst et al. 2018), fills critical gaps in our ability to model exoplanet atmospheres and provides useful laboratory data for interpreting and guiding current and future observations of super-Earths and mini-Neptunes, especially for TESS, JWST, and

WFIRST.

This work was supported by the NASA Exoplanets Research Program Grant NNX16AB45G. C.H. was supported by the Morton K. and Jane Blaustein Foundation.